\documentclass[twocolumn,reprint,superscriptaddress]{revtex4}
\usepackage{bm}
\usepackage{amssymb}
\usepackage{amsfonts}
\usepackage{graphicx}
\usepackage{bm}
\usepackage{subfigure}
\usepackage{multirow}

\begin{document}

\title{Estimation of Coupling Constants of a Three-Spin Chain: \\
Case Study of Hamiltonian Tomography with NMR}

\author{Elham Hosseini Lapasar}
\email{hosseini@alice.math.kindai.ac.jp}
\affiliation{Research Center for Quantum Computing,
Interdisciplinary Graduate School of Science and Engineering,
Kinki University, 3-4-1 Kowakae, Higashi-Osaka, 577-8502, Japan}
\author{Koji Maruyama}
\email{maruyama@sci.osaka-cu.ac.jp}
\affiliation{Department of Chemistry and Materials Science,
Osaka City University, Sumiyoshi, Osaka 558-8585, Japan}
\author{Daniel Burgarth}
\email{daniel@burgarth.de}
\affiliation{Institute of Mathematics and Physics, 
Aberystwyth University, Aberystwyth SY23 3BZ, United Kingdom}
\author{Takeji Takui}
\email{takui@sci.osaka-cu.ac.jp}
\affiliation{Department of Chemistry and Materials Science,
Osaka City University, Sumiyoshi, Osaka 558-8585, Japan}
\author{Yasushi Kondo}
\email{ykondo@kindai.ac.jp}
\affiliation{Research Center for Quantum Computing,
Interdisciplinary Graduate School of Science and Engineering,
Kinki University, 3-4-1 Kowakae, Higashi-Osaka, 577-8502, Japan}
\affiliation{Department of Physics, Kinki University, 3-4-1 Kowakae, Higashi-Osaka, 577-8502, Japan}
\author{Mikio Nakahara}
\email{nakahara@math.kindai.ac.jp}
\affiliation{Research Center for Quantum Computing,
Interdisciplinary Graduate School of Science and Engineering,
Kinki University, 3-4-1 Kowakae, Higashi-Osaka, 577-8502, Japan}
\affiliation{Department of Physics, Kinki University, 3-4-1 Kowakae, Higashi-Osaka, 577-8502, Japan}

\begin{abstract}

It has been shown that inter-spin interaction strengths in a spins-1/2 chain
 can be evaluated by accessing one of the edge spins only.
We demonstrate this experimentally for the simplest case, a three-spin chain, with nuclear magnetic resonance (NMR) technique. 
The three spins in the chain interact through nearest-neighbor Ising
 interactions under site-dependent transverse fields. 
The employed molecule is an alanine containing three $^{13}$C nuclei, each of which has spin-1/2.

\end{abstract}

\keywords{Hamiltonian tomography, NMR, Thermal state}

\maketitle

\section{Introduction}

Fabricating a quantum system that would perfectly function as we desire is 
very challenging. 
Even with the most advanced nanotechnology it is still difficult 
to build a structure that would have the exact values of system
parameters to make it work as we initially designed. 
This gives rise to the crucial necessity of system identification. 
In the context of quantum control, the system identification primarily
refers to the identification of the system Hamiltonian, 
also known as Hamiltonian tomography.

Yet, it is in general formidably hard to estimate the Hamiltonian: 
the number of necessary initial settings and measurements grows
exponentially as the system size becomes larger. 
To make the problem of Hamiltonian estimation more feasible, 
various schemes to reduce the complexity and/or to minimize the effect of
physical noise have recently been studied quite intensively.
Examples are indirect, but efficient, schemes of Hamiltonian tomography
of spin systems under limited access  
\cite{burgarth09a, burgarth09b, difranco09, wiesniak10, burgarth11},
and also an application of (classical) compressed sensing to the
quantum setting \cite{alireza11a, alireza11b}
that greatly reduces the overall complexity.

In the case of NMR, the interactions among spins in a molecule are usually determined by measuring
all spins at once. 
But what if we are allowed to access only a single spin to reconstruct 
the whole Hamiltonian as in the above examples? 
In this paper, we report such an indirect Hamiltonian tomography of a three nuclei 
spin system with NMR as a minimum model.
In the molecule, the spins effectively form a one-dimensional (1D) chain
with nearest-neighbor interactions, which are of the Ising type. 
We attempt to estimate the coupling constants by accessing solely the
end spin pretending as if we had no knowledge about the interactions
in advance. 
Then, we will make a comparison between the estimated coupling constants
and the known values as estimated by standard methods.

The analysis that we carry out is different from that presented in \cite{burgarth11,fasihi11}, 
where the coupling constants are estimated from the energy eigenvalues obtained
from the spectral peaks, which are obtained
from long time evolution of the spin at an end of the chain.
In contrast,
we obtain the coupling constants by fitting the time domain data without 
calculating the spectra in the present work. 
The materials presented in this paper therefore provide the first step toward the full verification of the
scheme discussed in \cite{burgarth11,fasihi11}. 
While data fitting is not computationally efficient for large systems, we find that it is very suitable for the three-spin
chain and more robust against relaxtion than \cite{burgarth11,fasihi11}. Our finding thus paves the way to indirect estimation of a Hamiltonian
of a small-scale noisy system where direct methods are not applicable.
In such circumstances, it is impossible to obtain data over longer 
time periods, which is fundamental for obtaining sharp spectral peaks.

\section{Theory}

In this section, we review how to estimate the spin-spin
interaction strengths in a three-spin Ising chain 
with site-dependent transverse fields. 
The model in our mind is a three homonucleus molecule, such as 
alanine with three $^{13}$C nuclei. We use liquid-state NMR to
control and measure the spins. 

The initial state is, thus, a thermal state 
\begin{eqnarray}
\rho_{\rm th}(T)=\frac{e^{-H_0/k_B T}}{{\mathrm{Tr}}[e^{-H_0/k_B T}]},
\label{Ham_t}
\end{eqnarray}
where
\begin{eqnarray*} 
H_0 = -\omega_{0} \left(I_{z}^{1}+I_{z}^{2}+I_{z}^{3} \right) 
\end{eqnarray*}
with $\displaystyle 
I_i^1 = \frac{\sigma_i}{2} \otimes I \otimes I, 
I_i^2 = I \otimes  \frac{\sigma_i}{2}\otimes I, 
I_i^3 = I \otimes I \otimes \frac{\sigma_i}{2}.$
Here $T$ is the temperature,   
$\omega_{0}$ is the common Larmor frequency of the spins, and 
$\sigma_i$ is the $i$th component of the Pauli matrices. 
We drop the interaction terms among spins and chemical shifts of the spins 
temporarily
since they are small enough compared with $\omega_{0}$. 
We note that Eq.~(\ref{Ham_t}) is defined in the laboratory frame. 
 
A weakly coupled system develops according to the Hamiltonian
\begin{eqnarray}
{\cal H}=\omega_{11} I_{x}^{1}+\omega_{12} I_{x}^{2}+
\omega_{13} I_{x}^{3}+J_{12} I_{z}^{1} I_{z}^{2}
+J_{23} I_{z}^{2} I_{z}^{3}.
\label{Ham}
\end{eqnarray}
Here, $\omega_{1i}$ 
and $J_{ij}$ characterize the transverse field of spin~$i$ and 
the coupling constant between spins~$i$ and $j$, respectively. 
We note that the Hamiltonian~(\ref{Ham}) is described in the rotating frames 
fixed to each spin.

Now we evaluate the dynamics of spin~1
\begin{equation}
M_k(t) \equiv
\langle I_k^1(t)\rangle=\mathrm{Tr}[\rho(t) I_k^1],
\end{equation}
where $k \in \{x, y, z\}$ and
\begin{equation}
\rho(t)=(e^{-i {\cal {H}}t})\rho_{\rm th}(T)({e^{-i {\cal {H}}t}})^{\dagger}
\end{equation}
is the density matrix of the system at time $t$.

\begin{figure}[htbp]
\begin{center}
{\includegraphics[scale=.70,clip]{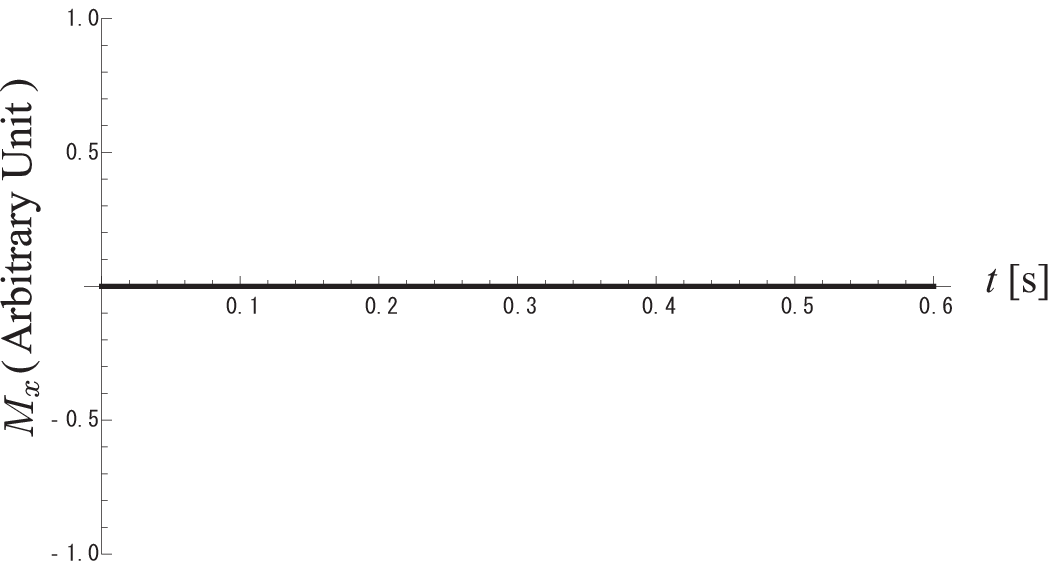}}
{\includegraphics[scale=.70,clip]{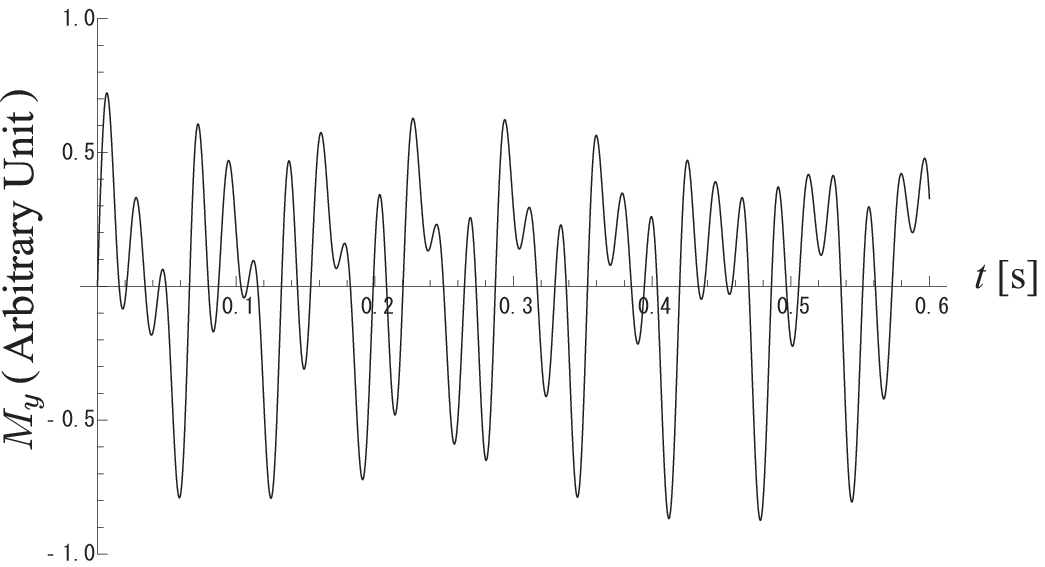}}
{\includegraphics[scale=.70,clip]{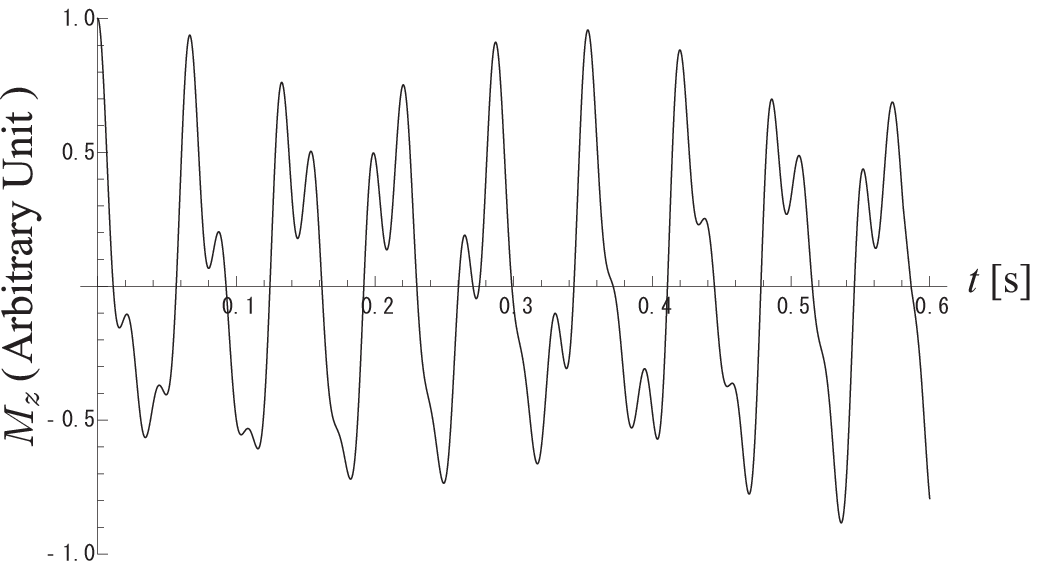}}
\caption{\label{fig:ideal}
Ideal dynamics of the expectation values $M_k(t)\ (k=x,y,z)$ of spin~1
when the initial state is a thermal state. 
Parameters
$\omega_{1i}/(2\pi)=27~{\rm Hz}$,
$J_{12}/(2\pi)=53.8~{\rm Hz}$, and
$J_{23}/(2\pi)=34.8~{\rm Hz}$ are employed in Eq.~(\ref{Ham}).} 
\end{center}
\end{figure}

The dynamics of spin~1 without 
relaxations nor transverse field inhomogeneities 
is calculated and shown in Fig.~\ref{fig:ideal} 
when $\omega_{1i}/(2\pi)=27~{\rm Hz}$ for all $i =1,2,3$. 
The coupling constants 
$J_{12}/(2\pi)=53.8~{\rm Hz}$ and 
$J_{23}/(2\pi)=34.8~{\rm Hz}$ 
are taken from \cite{kondo} as an example. 

It is clear that Fig.~\ref{fig:ideal} is far from reality. 
We then include the effects of transverse relaxations 
via the operator sum representation 
$\varepsilon(\rho)=\sum_{i=0}^{3}E_i^{\dagger}\rho E_i$ 
where $\sum_{i=0}^3 E_i^{\dagger} E_i=I$ \cite{Kraus, Jones}. 
We take 
\begin{eqnarray*}
E_0&=&\sqrt{\lambda_0}\,\,I, \\
E_{i}&=&\sqrt{1-\lambda_{i}} \,\, (2I_z^i) \quad (i=1,2,3),
\end{eqnarray*}
where
\begin{eqnarray*}
\lambda_0&=&\frac{1}{2}(-1+e^{-t/T_2(1)}+e^{-t/T_2(2)}+e^{-t/T_2(3)}),\\
\lambda_{i}&=&\frac{1}{2}(1+e^{-t/T_2(i)}) \quad (i=1,2,3),
\end{eqnarray*}
to represent the transverse relaxations. 

Finally, we take into account the inhomogeneity of the
transverse fields $\omega_{1i}(x)$ as a function of position $x$.  
We assume that the inhomogeneity has a Gaussian distribution 
\begin{eqnarray}
P(\omega_{1i}(x))
=\frac{1}{\sqrt{2\pi}\sigma} 
\exp\left[-\frac{(\omega_{1i}(x)-\omega_{1i})^2}{2\sigma^2}\right]
\label{eq:dist}
\end{eqnarray}
with variance $\sigma$ to be determined later. 

The coupling constants are estimated by comparing the spin dynamics
obtained numerically with various pairs $(J_{12}, J_{23})$
with the experimental data. Both the effects of the relaxations and the 
field inhomogeneity are taken into account in the numerical calculation.

\section{Experiment}

\label{sec:exp}
We employ a linearly aligned three-spin molecule for demonstrating 
a Hamiltonian tomography through an edge spin. 
Our task is to determine the scalar coupling
constants $J_{12}$ between spins~1 and 2 and $J_{23}$ between 
spins~2 and 3 by measuring only spin~1.

\subsection{Sample and Spectrometer}
\label{sec:spectrometer}

We demonstrate a Hamiltonian tomography of a spin system with  NMR.
We employ a JEOL ECA-500 NMR spectrometer \cite{jeol}, whose hydrogen
Larmor frequency is approximately 500~MHz.
We apply weak rf-fields to generate transverse fields in the
rotating frame of each spin.

A 0.3~ml, 0.78~M sample of ${}^{13}$C-labeled L-alanine 
(98\% purity, Cambridge Isotope) 
solved in D$_2$O, capsulated in a susceptibility matched NMR test 
tube \cite{t_tube}, is used.
Three ${}^{13}$C atoms are linearly aligned in L-alanine. 
We label  the carboxyl carbon  spin~1, the $\alpha$ carbon spin~2, 
and the methyl carbon spin~3.

The scalar coupling constants are estimated from the 
spectrum obtained by Fourier transforming the free induction decay (FID) 
signal after a hard $\pi/2$-pulse is applied for readout \cite{kondo}.
Here, protons are decoupled using a standard heteronucleus decoupling 
technique (WALTZ-16) \cite{freeman}.
The information extracted from the spectrum is summarized as follows.
The Larmor frequency differences are
$(\omega_{02}-\omega_{01})/2\pi=15.8$~kHz and
$(\omega_{03}-\omega_{02})/2\pi=4.4$~kHz,
where $\omega_{0i}$ denotes the Larmor frequency of the spin~$i$,
for which the chemical shift is considered. Large differences in the
Larmor frequencies compared with the scalar coupling constants justify
the weak-coupling assumption made when the Hamiltonian (\ref{Ham}) is
introduced. 
The scalar coupling constant $J_{13}$ between spins~1 and 3
is on the order of 1~Hz \cite{J13}, which is much
smaller compared to $J_{12}$ and $J_{23}$, and hence we can
safely ignore it in our analysis.
As a result,
the Hamiltonian of alanine molecule is well approximated by
Eq.~(\ref{Ham}). 
 
Measured relaxation times are 
$T_1(1)=15.5$~s, $T_1(2)=1.4$~s, $T_1(3)=0.9$~s and 
$T_2(1)=0.45$~s, $T_2(2)=0.23$~s, $T_2(3)=0.63$~s,
where the argument labels the spin. The spin~2 has the shortest $T_2$. 
In view of the fact that our data aquisition time to 
estimete the Hamiltonian is much shorter than any of $T_1(i)$, 
we ignore the effect of $T_1$ from now on. In contrast, we fully
take the effect of $T_2(i)$ into account in our numerical calculations.

The transverse fields are applied to spins 
by feeding oscillating currents with three different 
frequencies, corresponding to the Larmor
frequencies of the spins, to the coil. 
Their strengths are characterized by $\omega_{1i}$.

\subsection{Transverse Field Calibration}
\label{sec:C}

\begin{figure}[htbp]
\begin{center}
{\includegraphics[scale=.82,clip]{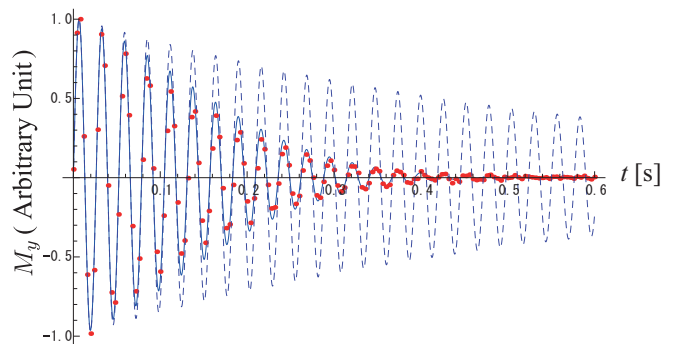}}
\caption{\label{fig:calibration}
(Color online) Calibration of strength and inhomogeneity of transverse field. 
Experimental results with a setting given in the text are shown. 
The dashed line shows the numerical result
in which only the effect of relaxation is taken into account,
while the solid line is the result in which both 
 relaxations and inhomogeneities of transverse fields are considered.
}
\end{center}
\end{figure}

We measure the dynamics of spin~1 in the presence of $\omega_{11}/(2\pi)
=27$~Hz only, while
other $\omega_{1i}\ (i=2,3)$ is set to 0, as shown in
Fig.~\ref{fig:calibration}. The data was acquired in every $0.004$~s for
$0 \leq t \leq 0.6$~s.
The periodicity provides the information on the strength of the transverse
fields, 
while the decay rate
is determined by the relaxations and the field inhomogeneities. 
We find that the relaxations only are not enough to reproduce the decay, 
as demonstrated by the dashed line in Fig.~\ref{fig:calibration}. 
Both relaxations and field inhomogeneities must be considered to reproduce 
the decay rate. 
We obtain the variance $\sigma/\omega_{1i} = 0.05 $ in Eq.~(\ref{eq:dist})
by fitting the data.

\subsection{Results}
\label{sec:R}
We measure the dynamics of spin~1 in two cases.

\begin{figure}[htbp]
\begin{center}
{\includegraphics[scale=.70,clip]{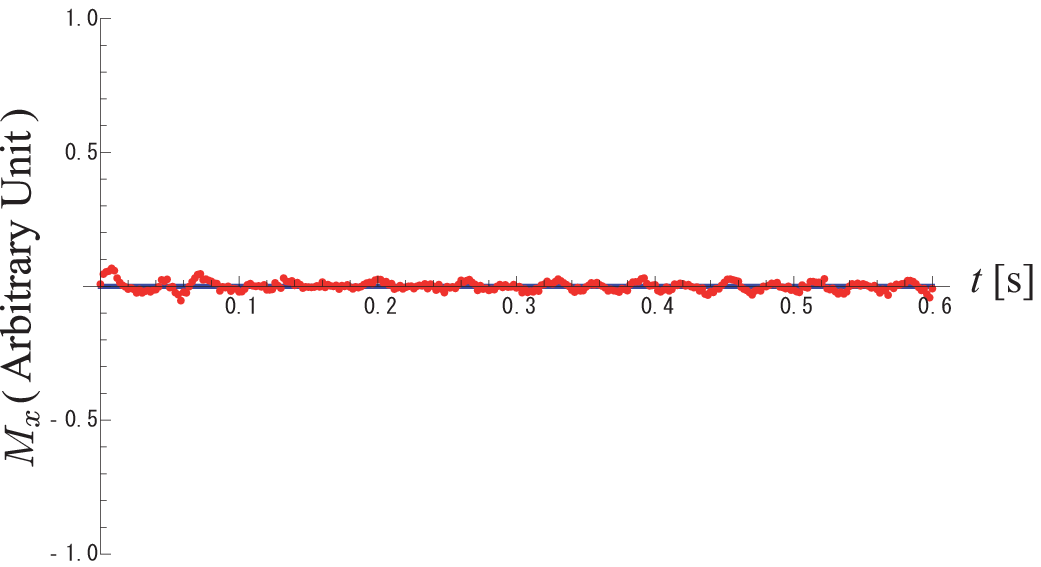}}
{\includegraphics[scale=.70,clip]{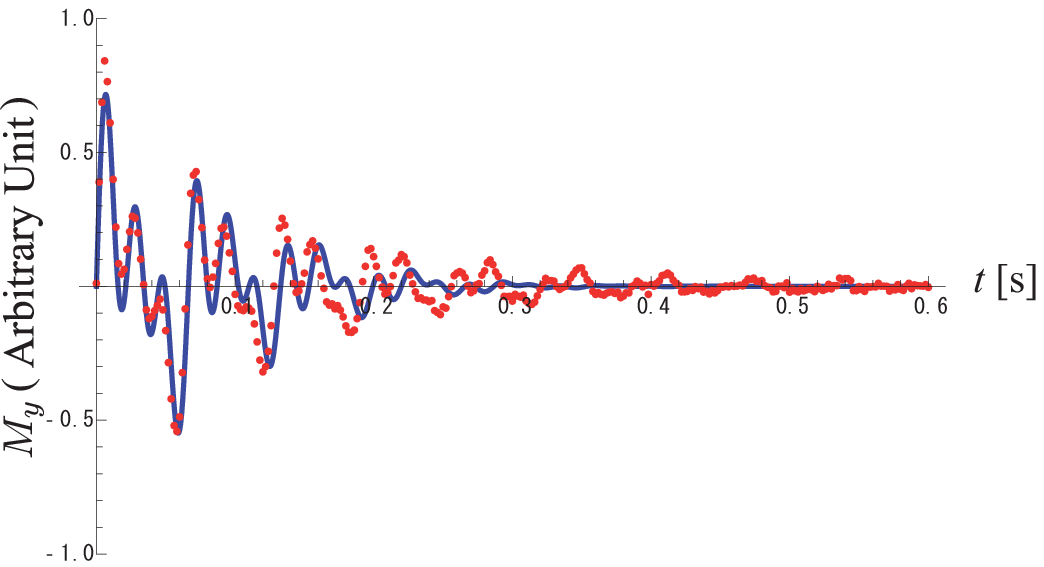}}
{\includegraphics[scale=.70,clip]{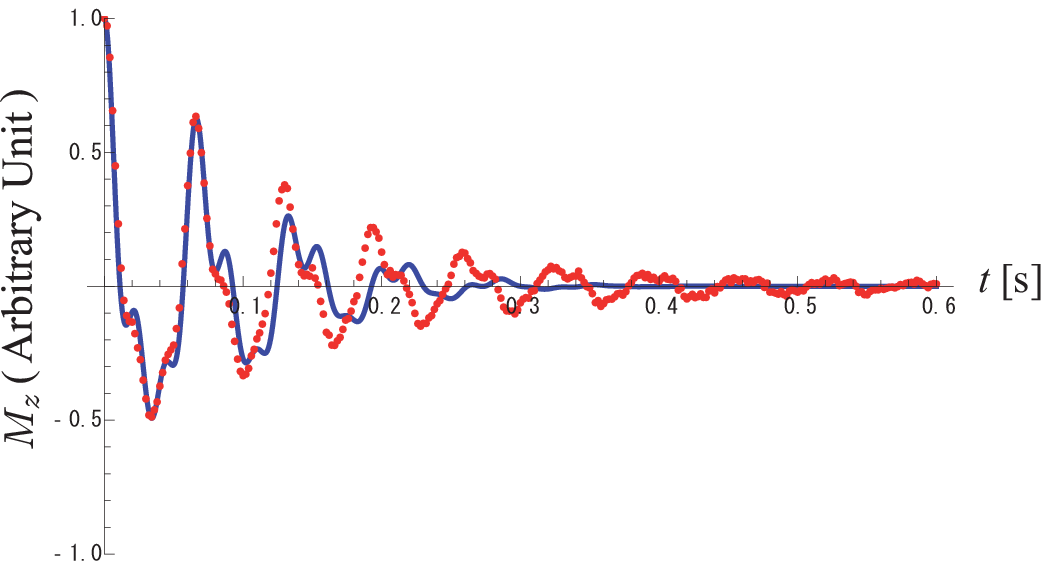}}
\caption{\label{fig:exp_1}
(Color online) 
Dynamics of $M_x, M_y,$ and $M_z$ of spin~1 
are shown for Case 1, when all $\omega_{1i}/(2\pi)= 27$~Hz and 
the initial states of all spins are thermal.
Experimental Results are shown by dots, and
the solid lines are the numerical results, in which known values of coupling constants are employed.
}
\end{center}
\end{figure}

In Case 1, the initial state is thermal and
the transverse fields $\omega_{1i}/(2\pi) =27$~Hz are applied to
all the spins.
The dynamics of the expectation values 
$M_x, M_y,$ and $M_z$ of spin~1 are shown in
Fig.~\ref{fig:exp_1} as functions of time $t$.
In Fig.~\ref{fig:exp_1}, we see there are structures different from a simple
sinusoidal oscillation which is expected without interactions. 
In other words, we obtain information concerning the interactions by 
measuring spin~1 only. It should be noted, however, that 
the spin dynamics is strongly affected by relaxations and 
transverse field inhomogeneities of all spins.

\begin{figure}[htbp]
\begin{center}
{\includegraphics[scale=.70,clip]{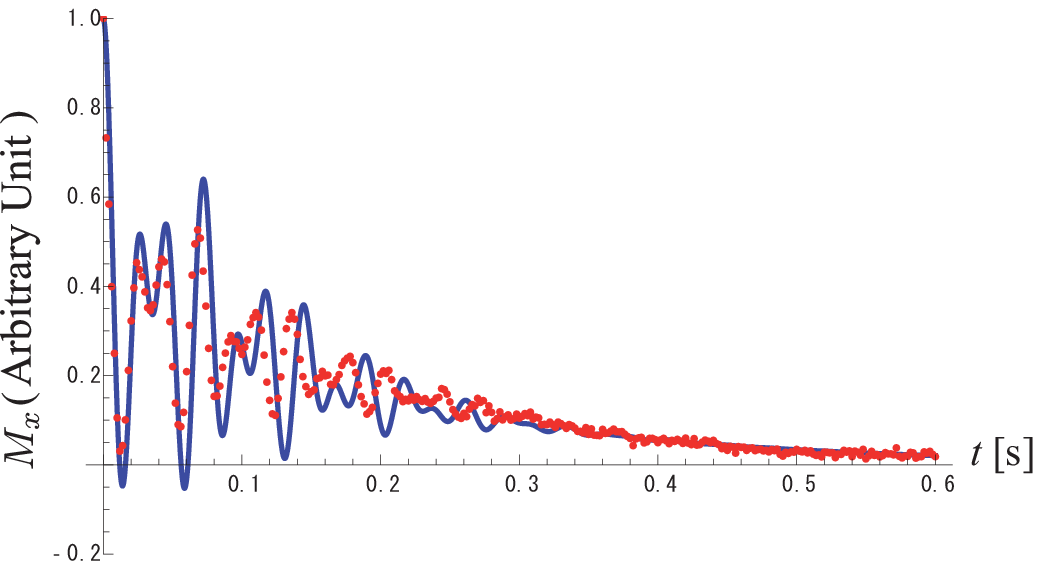}}
{\includegraphics[scale=.70,clip]{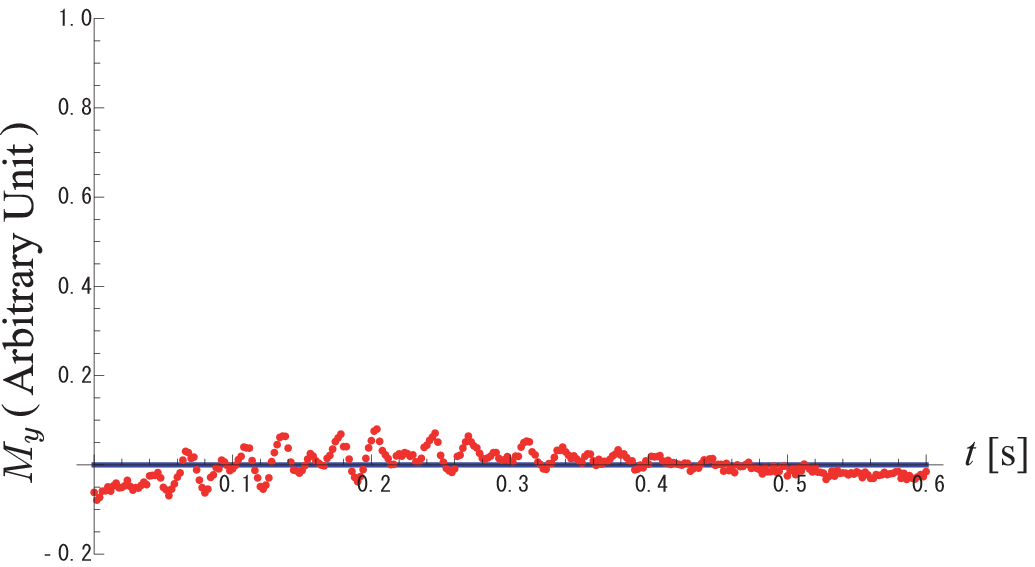}}
{\includegraphics[scale=.70,clip]{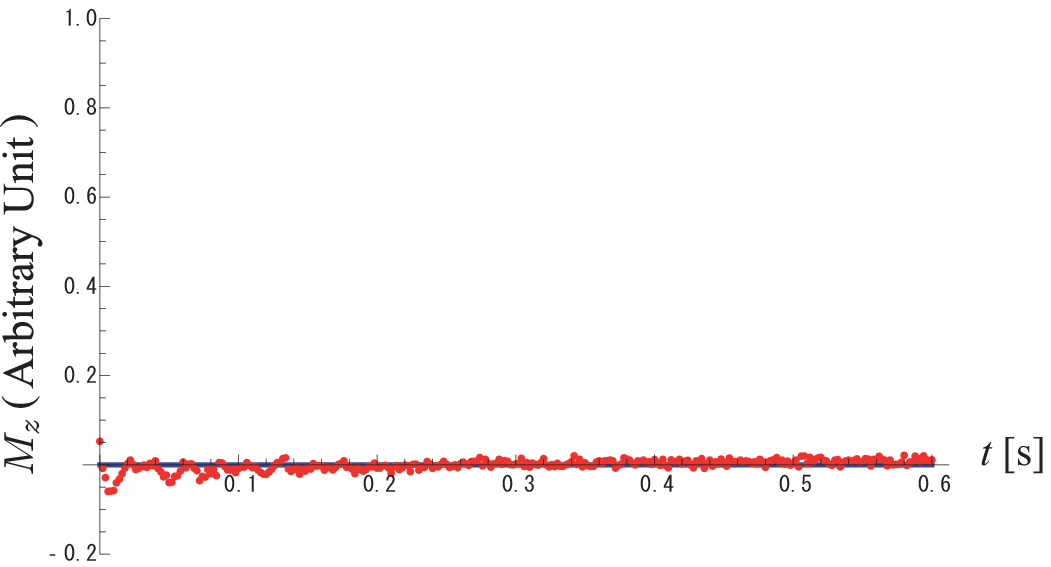}}
\caption{\label{fig:exptt_2}
(Color online)
Dynamics of $M_x, M_y,$ and $M_z$ of spin~1 
are shown for Case 2, when
$\omega_{11}= 0$ and $\omega_{12}/(2\pi)=\omega_{13}/(2\pi)=27$~Hz.
The initial state of spin~1 is prepared by applying a $\pi/2$-pulse 
along the $y$-axis to the thermal state, 
while those of spin~2 and 3 remain thermal.
Experimental results are shown by dots, and
the solid lines show the numerical results, in which known values of coupling constants are used.
}
\end{center}
\end{figure}

In Case 2, we have chosen
$\omega_{11} = 0$, $ \omega_{12}/(2\pi)=\omega_{13}/(2\pi)=27$~Hz. 
Since $\omega_{11}=0$, we would not expect any dynamics in spin 1 if a thermal 
state were employed as an initial state. To avoid this problem,
the initial state of spin~1 is prepared by
applying a $\pi/2$-pulse along the $y$-axis to the thermal 
equilibrium state, while
the initial states of spins~2 and 3 remain thermal. 
We again obtain the information on the interactions by measuring
only spin~1 as shown in Fig.~\ref{fig:exptt_2}.

\section{Estimation of Coupling Constants}

In this section, we pretend as if we do not know the coupling
constants $J_{12}$ and $J_{23}$ and, instead,
we estimate them by fitting numerically evaluated
$\langle I_k^1(t) \rangle$ with various 
values of ($J_{12}, J_{23}$) to the experimental data.
Although we defined the magnetization $M_k(t)$ of the
first spin as the expectation value $\langle I_k^1(t)\rangle$,
we temporarily assign $M_k(t)$ to the experimental data while
$\langle I_k^1(t) \rangle$ to the corresponding numerical result to avoid
confusion.
Let us define the ``distance'' between the
experimental data $M_k(t)$ and the 
numerical result $\langle I_k^1(t)\rangle$ by
\begin{eqnarray*}
D_k(J_{12},J_{23})
= \sqrt{\sum_{j} |\langle I_k^1(t_j,J_{12},J_{23}) \rangle - M_k(t_j)|^2}.
\end{eqnarray*}
Here $\{t_j\}$ denotes the set of data acquisition points and
$\langle I_k^1(t_j,J_{12},J_{23}) \rangle$
is the numerically evaluated expectation value 
of the $k$th component of spin 1
at time $t_j$ with the coupling constants $(J_{12},J_{23})$.
In actual experiment, data was acquired in every $0.002$~s for $0 \leq t
\leq 0.6$~s.

We do not make use of the spectra obtained by the Fourier transforms of 
time domain signals since the time window is not large enough to provide
sharp peaks in the spectra. We can freely select the fitting window
from $t=0$ to $t_{\rm w}> t_0$, where
$\displaystyle t_0 = 2\pi/J_{12}+2\pi/J_{23} \sim
50$~ms is the minimum time required
for information to propagate from spin~1 to spin~3 through
spin~2 and then propagate back to spin~1. 
Clearly there is an optimal value for $t_{\rm w}$, 
since too small $t_{\rm w}$ provides
too little data to be fitted, while too large $t_{\rm w}$ makes
relaxations and field inhomogeneities too significant. 
We estimate the coupling constants for three different values of
$t_{\rm w}$ and compare the results in the following.

\subsection{Case 1: $\omega_{11}\neq 0$}

In this case, the initial states of the three spins are prepared in
thermal states and transverse fields $\omega_{1i}/(2\pi)=27~{\rm Hz}$
$(i=1,2,3)$ are applied to all three spins. 

\begin{figure}[h]
\begin{center}
{\includegraphics[scale=.65,clip]{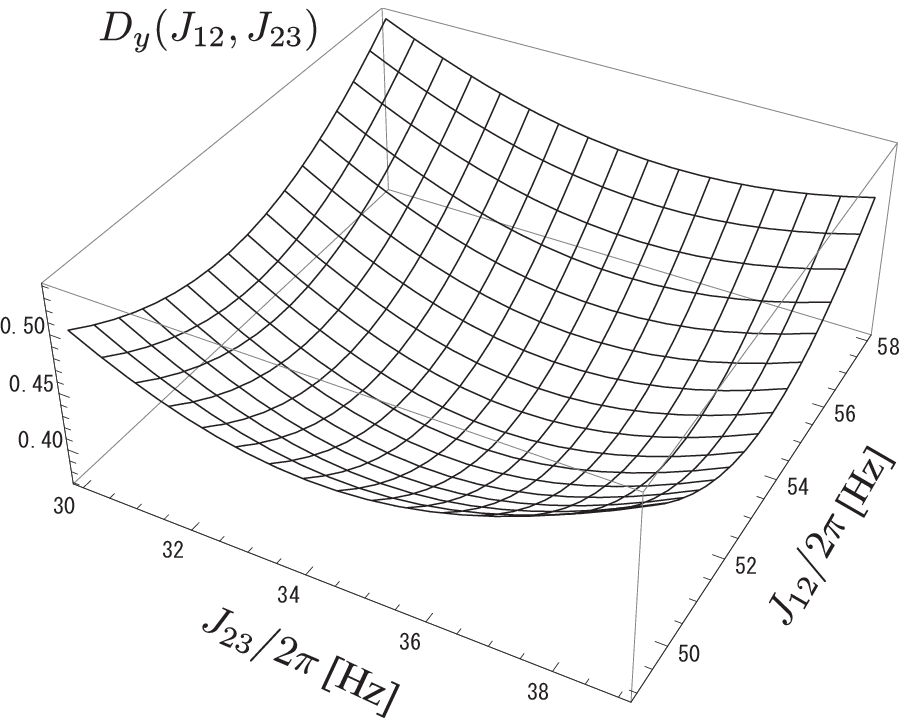}}

\vspace{1cm}

{\includegraphics[scale=.55,clip]{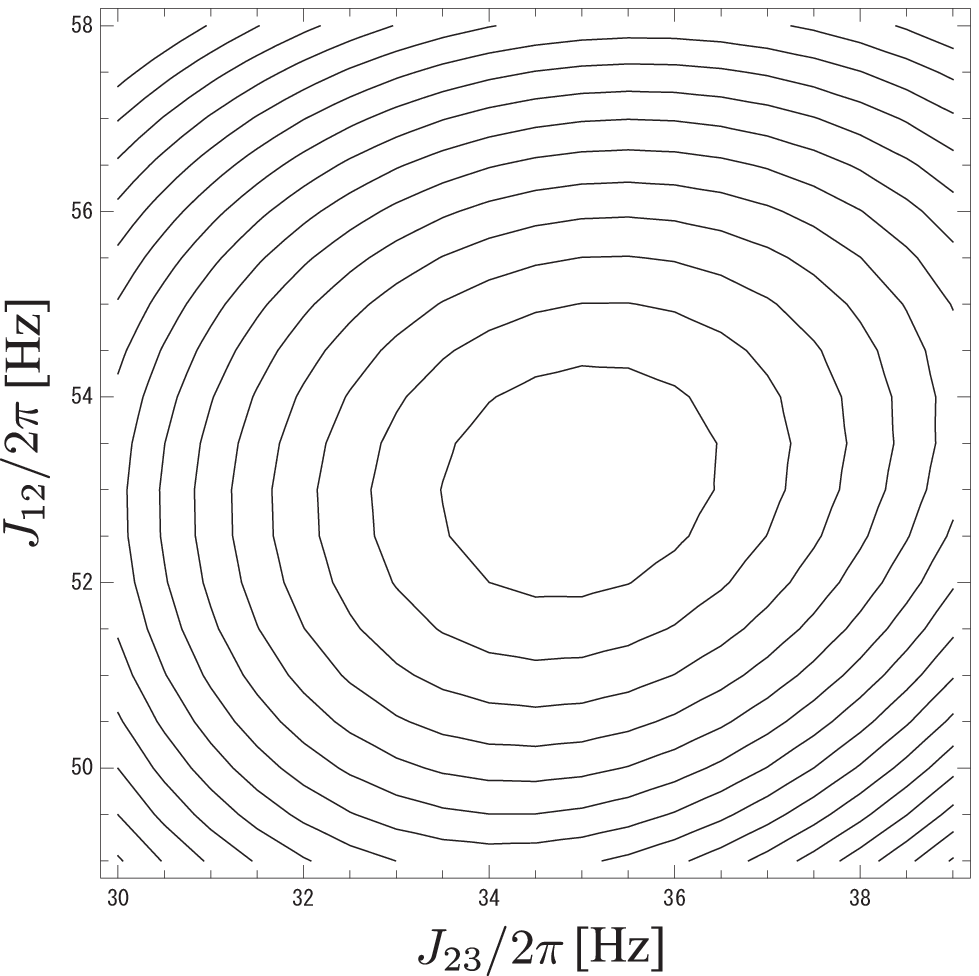}}
\caption{\label{fig:1y}
3D and contour plot of $D_y(J_{12}, J_{23})$ for Case 1 when $t_{\rm
 w}=0.05$~s. A clear minimum can be seen. The distance between
two neighboring contours in the contour plot is 0.01.
}
\end{center}
\end{figure}

\begin{figure}[h]
\begin{center}
{\includegraphics[scale=.65,clip]{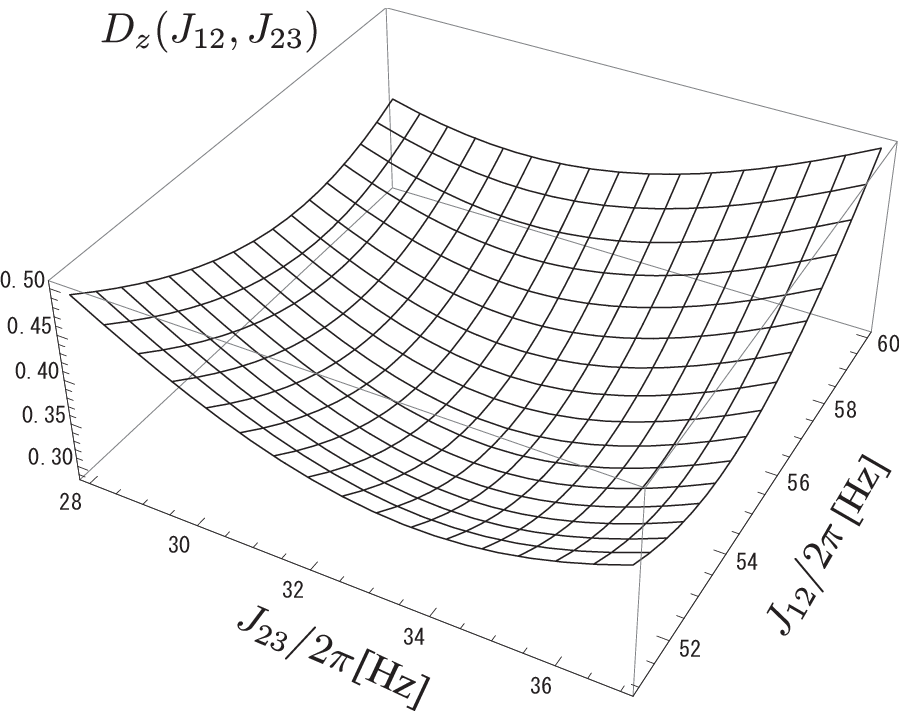}}

\vspace{1cm}

{\includegraphics[scale=.55,clip]{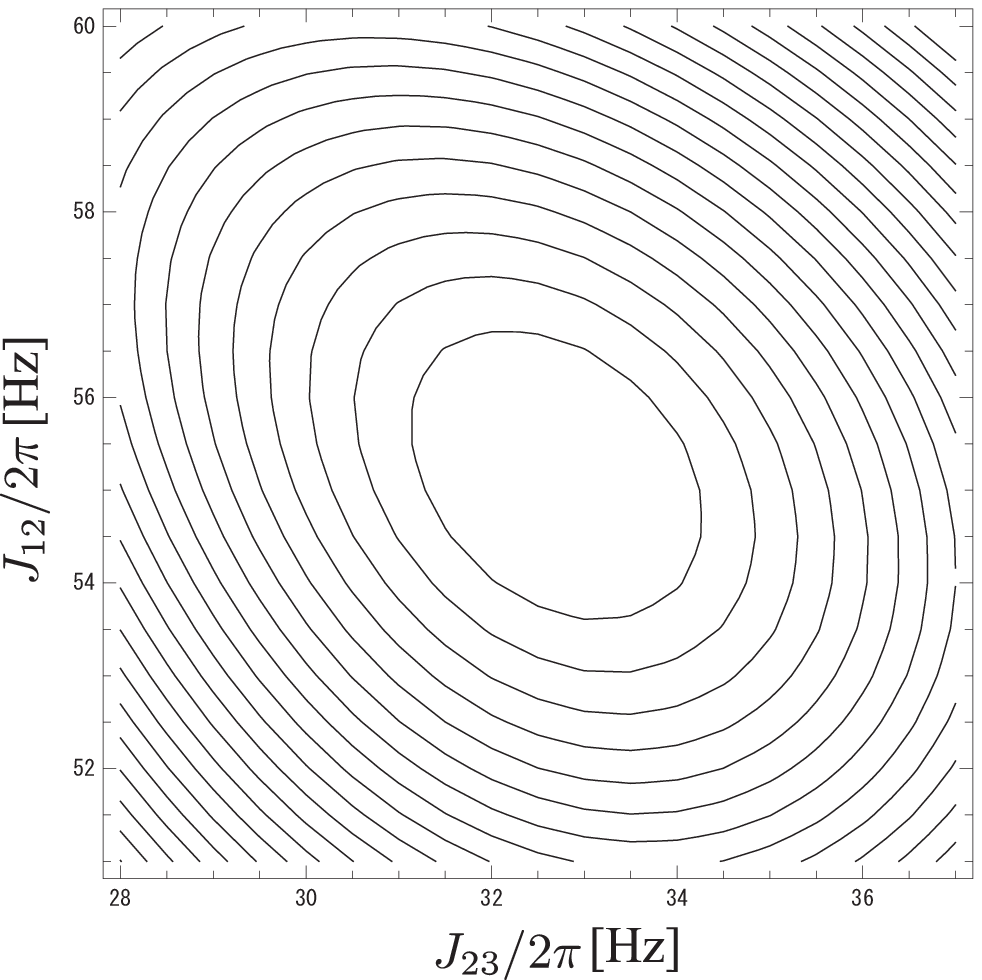}}
\caption{\label{fig:1z}
3D and contour plot of $D_z(J_{12}, J_{23})$ for Case 1 when $t_{\rm
 w}=0.05$~s. A clear minimum can be seen. The distance between
two neighboring contours in the contour plot is 0.01.
}
\end{center}
\end{figure}

3D and contour plots of the distances
$D_y(J_{12}, J_{23})$ and $D_z(J_{12}, J_{23})$ with $t_{\rm w} =
0.05$~s are shown in Fig.~\ref{fig:1y} and Fig.~\ref{fig:1z}, respectively. 
As expected, clear minima are found in these plots. 

We obtain the set $(J_{12}, J_{23})$ that minimizes
$D_y(J_{12}, J_{23})$ and $D_z(J_{12}, J_{23})$ for
different $t_{\rm w} = 0.05, 0.1, 0.2$~s. 
They are summarized in Table \ref{tab:1}.

\begin{table}[h]
\caption{Estimated coupling constants in Case 1 for various window size 
$t_{\rm w}$.}
\centering
\label{tab:1}
\begin{tabular}{c c c c}
\hline\hline
Case 1& $t_{\rm w}$~[s]~~ & $J_{12}/2\pi$~[Hz]~~ & $J_{23}/2\pi$~[Hz]
\\ [0.5ex]
\hline
Known values & & $53.8$ & $34.8$\\
\hline
\multirow{3}{*}{$D_y$}
 & $0.05$ & $53$ & $35$\\
 & $0.1$ & $55.5$ & $36$ \\
 & $0.2$ & $55.5$ & $36.5$ \\
\hline 
\multirow{3}{*}{$D_z$} 
 & $0.05$ & $55$ & $33$\\
 & $0.1$ & $54.5$ & $30.5$ \\
& $0.2$ & $56.5$ & $25.5$ \\

\hline  
\end{tabular}
\end{table}

\subsection{Case 2: $\omega_{11}=0$}

We take $\omega_{11}=0$ and $\omega_{12}/(2\pi)=\omega_{13}/(2\pi)
=27~{\rm Hz}$ in
Case 2. To introduce nontrivial spin dynamics to spin~1,  
a $\pi/2$-pulse along the $y$-axis, $Y=\exp(-i\pi I_y^1/2)$, is
applied to spin~1 at $t=0$ after the thermal state has been prepared.

\begin{figure}[h]
\begin{center}
{\includegraphics[scale=.65,clip]{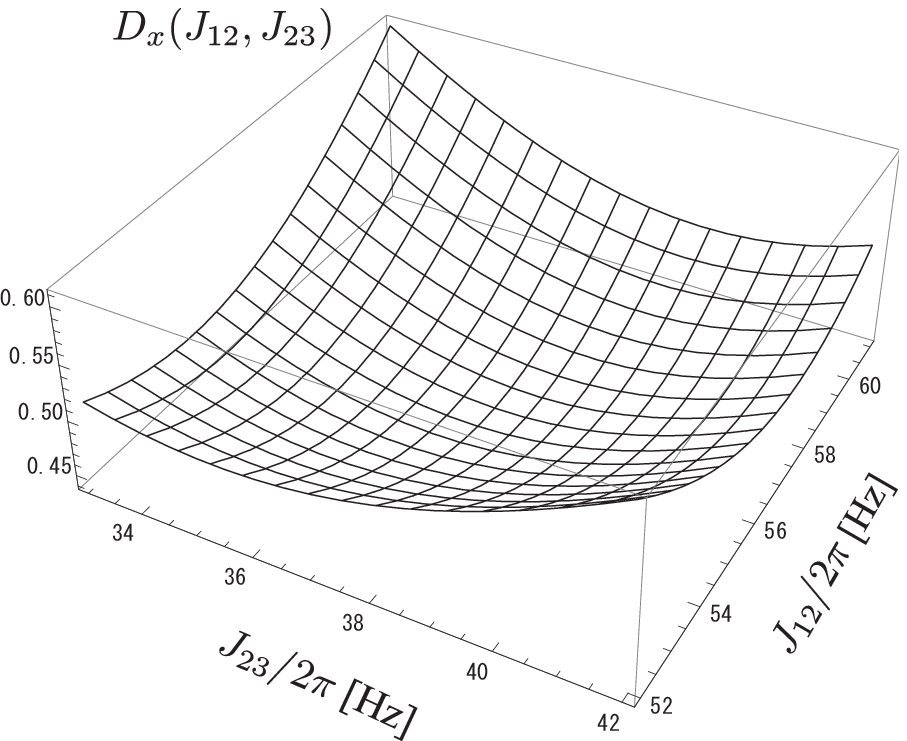}}

\vspace{1cm}

{\includegraphics[scale=.55,clip]{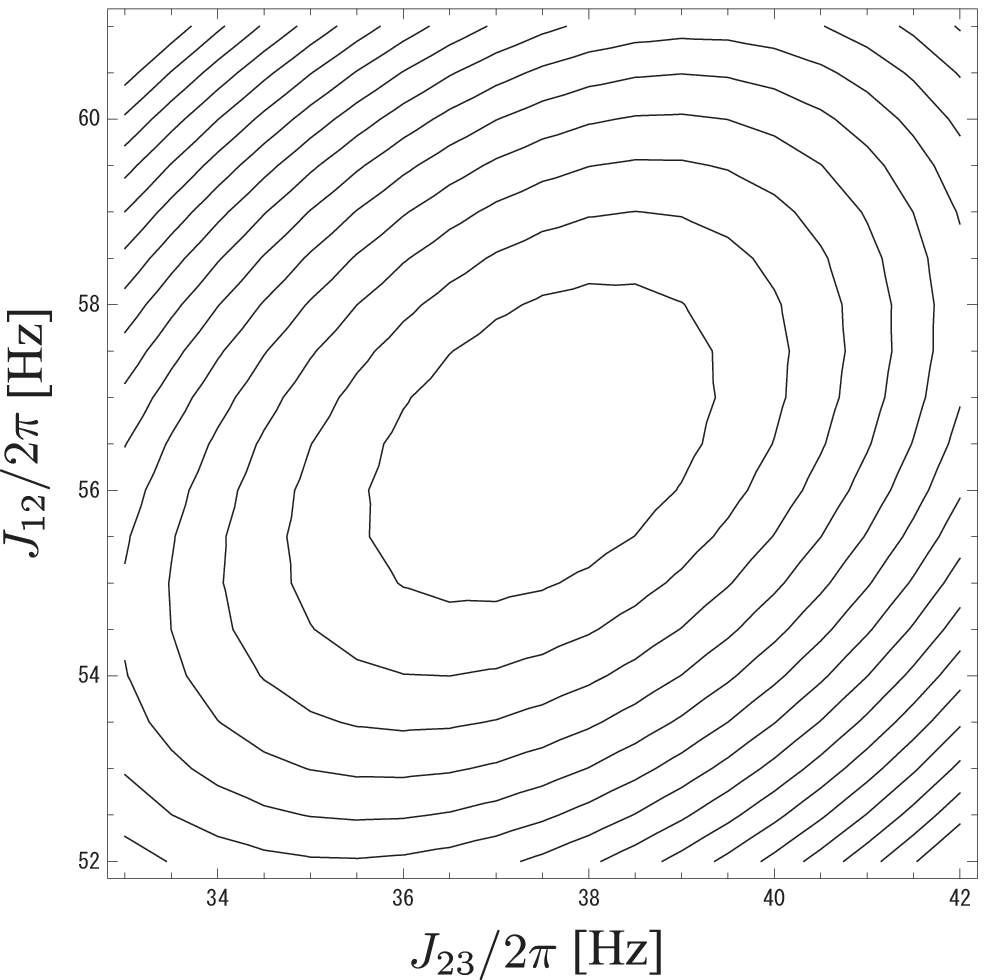}}
\caption{\label{fig:2x}
3D and contour plots of $D_x(J_{12}, J_{23})$ for Case 2 with time window
$t_{\rm w}=0.05$~s. A clear minimum can be seen. The distance between
two neighboring contours in the contour plot is 0.01.
}
\end{center}
\end{figure}

Figure~\ref{fig:2x} shows the 3D and contour plots of the distance
$D_x(J_{12}, J_{23})$ with $t_{\rm w} =0.05$~s.
As expected, a unique minimum can be found in the figure. 

We obtain the pair $(J_{12}, J_{23})$ that minimizes the distance
$D_x(J_{12}, J_{23})$ with different time windows
$t_{\rm w} = 0.05, 0.1, 0.2$~s. 
Table \ref{tab:2} summarizes the results.

\begin{table}[h]
\caption{Estimated coupling constants in Case 2 for different 
window size
$t_{\rm w}$.}
\centering
\label{tab:2}
\begin{tabular}{c c c c}
\hline\hline
Case 2~~~ & $t_{\rm w}$~[s]~~ & $J_{12}/2\pi$~[Hz]~~ & $J_{23}/2\pi$~[Hz]
\\ [0.5ex]
\hline
Known values & & $53.8$ & $34.8$\\
\hline
\multirow{3}{*}{$D_x$} 
 & $0.05$ & $56.5$ & $37.5$\\
 & $0.1$ & $58$ & $38$ \\
& $0.2$ & $59$ & $38.5$ \\
\hline 
\end{tabular}
\end{table}

\begin{figure}[t]
\begin{center}
{\includegraphics[scale=.65,clip]{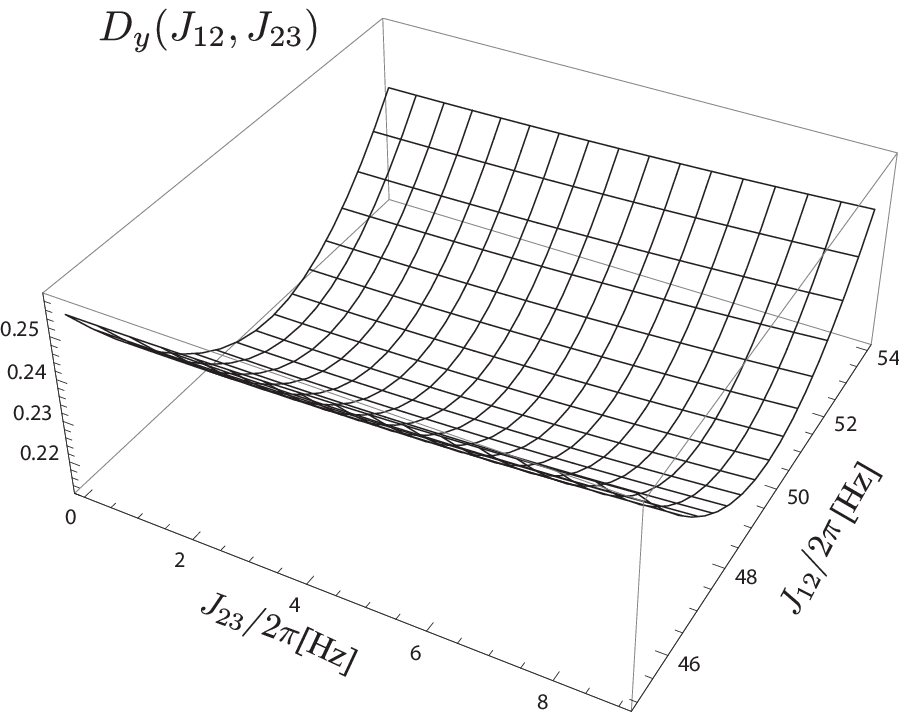}}

\vspace{1cm}

{\includegraphics[scale=.55,clip]{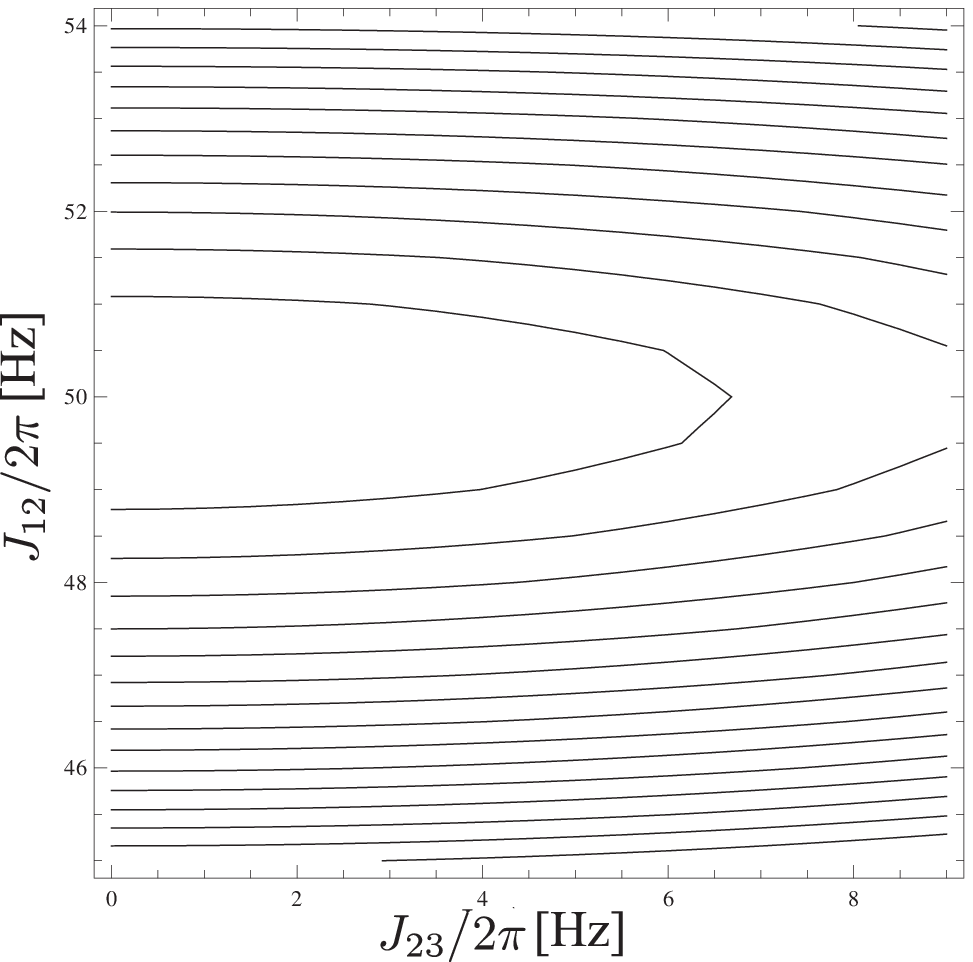}}
\caption{\label{fig:0.02}
3D and contour plots of $D_y(J_{12}, J_{23})$ for Case 1 with time window
$t_{\rm w}=0.02$~s. It shows that the time is not large enough to estimate the coupling constant $J_{23}$. The distance between
two neighboring contours in the contour plot is 0.0028.
}
\end{center}
\end{figure}

\subsection{Estimation}

Regardless of the choice of 
$t_{\rm w}$ or ($D_x$, $D_y$, $D_z$), the estimated pair $(J_{12}, J_{23})$
is consistent with each other both in Cases 1 and 2. It seems, 
however,
that the smallest $t_{\rm w}=0.05$~s yields the best results when we compare 
them with the known $(J_{12}, J_{23})$ obtained by different means. 
We have also confirmed that a smaller value, $t_{\rm w}=0.02$~s, is not 
large enough to estimate the coupling constants reliably as shown in
Fig.~\ref{fig:0.02}, where the profile has a sharp minium along 
the $J_{12}$-direction but is almost flat along the $J_{23}$-direction.
This bahavior clearly shows the significance of the time $t_0 \sim 
0.05$~s defined previously. The effect of $J_{23}$ does not manifest itself
yet
in the behavior of spin 1 for a short time less than $t_0$.
On the other hand, for $t_{\rm w} \sim t_0$, 
relaxation and field inhomogeneity are less serious yet, and the data
produces an excellent result, 
while the results provided by a larger $t_{\rm w}$
suffer from these effects.

\section{Summary}

We have successfully demonstrated for the first time that indirect Hamiltonian tomography is possible in NMR setup.
As long as the system is small enough for efficient data fitting, the estimated values are surprisingly close to
the real ones, given the substantial amount of noise and inhomogeneities in the system.
This paves the way towards the identification of spins and couplings which are off-resonant or would usually 
the drowned by background noise. While the methods of \cite{burgarth11,fasihi11} rely on Fourier 
analysis, which is only applicable in systems clean enough for sufficiently long time data aquisition, our method can be applied in more noisy cases.

We have shown that there is a competition in the observed evolution between amount of data acquired and the amount of noise coming in.
It seems optimal to choose rather short data acquisition times in order to get a good agreement of the estimated couplings with their real 
values.

\section*{Acknowledgments}

The work of EHL, YK and MN is supported by 
\lq Open Research Center\rq~Project for Private Universities; matching 
fund subsidy from MEXT (Ministry of Education, Culture, Sports, Science 
and Technology). YK and MN would like to thank partial supports 
of Grants-in-Aid for Scientific Research from the JSPS (Grant No.~23540470).
KM is grateful to the support by the JSPS Kakenhi (C) (Grant No.~22540405).
KM and TT are supported in part by Quantum Cybernetics (Grant No.
2112004), CREST-JST, and FIRST-JSPS (Quantum Information Process).

\end{document}